# Cohesive energy and structural parameters of binary oxides of groups IIA and IIIB from diffusion quantum Monte Carlo

Juan A. Santana,[1,2] Jaron T. Krogel,[1] Paul R. C. Kent,[3,4] Fernando A. Reboredo[1,a]

[1] Materials Science and Technology Division, Oak Ridge National Laboratory, Oak Ridge, Tennessee 37831, USA

[2] Department of Chemistry, University of Puerto Rico at Cayey, P. O. Box 372230, Cayey, PR 00737-2230, USA

[3] Center for Nanophase Materials Sciences, Oak Ridge National Laboratory, Oak Ridge, Tennessee 37831, USA

[4] Computer Science and Mathematics Division, Oak Ridge National Laboratory, Oak Ridge, Tennessee 37831, USA

**ABSTRACT:** We have applied the diffusion quantum Monte Carlo (DMC) method to calculate the cohesive energy and the structural parameters of the binary oxides CaO, SrO, BaO, $Sc_2O_3$, $Y_2O_3$ and $La_2O_3$. The aim of our calculations is to systematically quantify the accuracy of the DMC method to study this type of metal oxides. The DMC results were compared with local, semi-local and hybrid Density Functional Theory (DFT) approximations as well as with experimental measurements. The DMC method yields cohesive energies for these oxides with a mean absolute deviation from experimental measurements of 0.18(2) eV, while with local, semi-local and hybrid DFT approximations the deviation is 3.06, 0.94 and 1.23 eV, respectively. For lattice constants, the mean absolute deviation in DMC, local, semi-local and hybrid DFT approximations, are 0.017(1), 0.07, 0.05 and 0.04 Å, respectively. DMC is highly accurate method, outperforming the DFT approximations in describing the cohesive energies and structural parameters of these binary oxides.

[a] Electronic mail: reboredofa@ornl.gov.

## I. INTRODUCTION

Diffusion quantum Monte Carlo (DMC) is a stochastic sampling method to solve the many-body Schrödinger equation.[1] It is a universal approach that captures all electron dynamics on an equal footing across atomic, molecular and condensed systems without empirical parameters. Such methods are crucial for reliable simulations of the complex heterogeneous nature of current materials, and even more importantly, for the prediction of new materials. These advantages have been recognized, and the application of DMC has increased significantly in the last few years;[2–19] see Ref. 20 for a review of applications before 2013.

As in any method going into the application stage, there are many open questions regarding the best protocols for calculations and the accuracies that can be achieved in practice. Even though DMC is formally an exact method, it requires various approximations for practical reasons.[1] The most significant approximations are the fixed-node (FN)[1] (or the generalized fixed-phase)[21] approximation, the use of pseudopotentials and the localization approximation.[22,23] Other approximations, that are simpler to control, include the short time-approximation and the use of supercells to simulate condensed matter.[1] For these approximations, running longer or larger calculations results in systematically smaller errors. Multiple methods are available to reduce the uncertainty introduced by these approximations and various studies[3,19] have systematically evaluated how the DMC results depend on them. For instance, the set of methods most commonly used in DMC calculations (denoted hereafter as the "standard DMC method", see details in Sec. II) was benchmarked in Ref. 3 for the volumes and bulk moduli of many different type of solids. The findings of Ref. 3 show that standard DMC is highly accurate in describing the structural properties of a broad range of solids. The results also indicate that structural properties of solids are rather insensitive to the nodal surface.[3]

In the present work, we applied standard DMC to calculate the cohesive energy and the structural parameters of a series of binary oxides: CaO, SrO, BaO, $Sc_2O_3$, $Y_2O_3$ and $La_2O_3$. In the spirit of Ref. 3, the aim of our calculations is to systematically quantify the accuracy of the standard DMC method to study this type of metal oxides. The DMC method yields cohesive energies and structural parameters for these oxides with a mean absolute deviation from experimental measurements of 0.18(2) eV and 0.017(1) Å, respectively. These results are in line with the benchmark calculations presented in Ref. 3, showing that DMC is highly accurate in describing the structural properties of a broad range of solids and that these structural properties are rather insensitive to the nodal surface. Our calculations for oxides of groups IIA and IIIB are one of the first efforts to systematically assess how accurately standard DMC methods can predict the cohesive energy of solids. The results show that accurate cohesive energies can already be obtained for simple oxides with standard DMC methods.

## II. METHODOLOGY

### A. DMC calculations and supercell models

DMC calculations were performed with QMCPACK[24] (http://qmcpack.org). Unless otherwise specified, we use standard methodologies[3] in our DMC calculations: *i*) fixed-node approximation with single determinant Slater-Jastrow trial wavefunction and single-particle orbitals generated within the Local Density Approximation (LDA),[25] *ii*) norm-conserving pseudopotentials (see details in Sec. B), and *iii*) many-body finite-size (FS) effects corrected with twist averaged boundary conditions[26] and the method proposed by Kwee, Zhang and Krakauer (KZK).[27] The Jastrow factor included one- and two-body terms with parameters optimized by variance minimization.[28] The single-particle orbitals were generated with the plane-wave based code Quantum ESPRESSO.[29] The plane-wave energy cutoff was set to 4082 eV (300 Ry). A small DMC time-step of 0.0025 Ha$^{-1}$ was used.[3]. The scheme proposed by Casula[23] (T-moves) was used to treat the nonlocal part of the pseudopotentials (PPs) within DMC and avoid numerical instabilities in the locality approximation.[22] Calculations for bulk phases were performed with the GPU implementation[30] of DMC in QMCPACK.[24] Employing NVIDIA Tesla K20X GPU accelerators, DMC calculations of the equation of state (EOS) of one of the studied oxides (Figure 2) costs ~160,000 compute hours. The computational cost is ~6 times higher on traditional processors (AMD Opteron 6274). The workflow automation system Nexus[31] was used to manage and monitor the various stages of the calculations.

The EOS of CaO, SrO and BaO were calculated for their known stable structures, i.e., rock salt (space group $Fm\bar{3}m$).[32–34] Similarly, $Sc_2O_3$, $Y_2O_3$ and $La_2O_3$ were simulated on their know stable structures,[35,36] i.e., space group $Ia\bar{3}$ for $Sc_2O_3$ and $Y_2O_3$ and $P\bar{3}m1$ for $La_2O_3$. The atomic positions were taken from the Inorganic Crystal Structure Database (ICSD):[37] $Sc_2O_3$ (code: 41264), $Y_2O_3$ (code: 41267) and $La_2O_3$ (code: 56771). For $La_2O_3$, which has a hexagonal structure, we evaluated the EOS approximately by fixing the *c/a* ratio to the experimental[36] value 1.5597. The binary oxides of group IIA were simulated with a 3×3×3 (54 atoms) supercell. Oxides of group IIB were simulated with 40 atom supercells. Calculations were performed with twist averaged boundary conditions[26] on a 2×2×2 supercell grid.

### B. Pseudopotential generation

Norm-conserving PPs were generated with OPIUM[38] for the O, Ca, Sr, Ba, Sc, Y and La atoms. The O-PP is based on a He-core PP. We previously[12] tested the O-PP within DMC by evaluating the ionization potential (IP) of oxygen and the equilibrium distances ($r_e$) and dissociation energies ($D_e$) of the $O_2$ dimer. The IP of O evaluated with DMC agrees with the experimental value to within 0.01 eV, and $D_e$ is within 0.15 eV of the experimental value.[12] Similarly, the DMC equilibrium distance $r_e$ of $O_2$ also agrees very well the with experiment value (within 0.01 Å). Further details of our O-PP can be found in Ref. 12. Below we describe the Ca-, Sr-, Ba-, Sc-, Y- and La-PPs, which have not been presented before. Pseudopotential files are available in the Supplemental Material[39] section in formats compatible with the Quantum ESPRESSO[29] and QMCPACK[24] packages.

There are various possible core-valence partitions to generate PP for atoms of group IIA and IIIB. For instance, the core can be taken as Ar, Ni or Kr for Sr, and as Kr, Pd, or Xe for La. However, for extended systems, the computational cost of small core PPs (e.g., Ar or Kr for Sr and La, respectively) will be excessively high. The large core PP (Kr or Xe), on the other hand, can be expected to yield inaccurate results. Therefore, PPs with 10 and 11 electrons (for instance, Ni-core and Pd-core PP for Sr and La, respectively), represent a compromise between accuracy and computational cost. Similar PPs have been previously developed for QMC calculations with a Gaussian basis set representation, e.g., for atoms of group IIA[40] and Sc.[41] However, our PPs are developed instead for compatibility with the plane wave representation. This representation is advantageous for periodic systems because it can be converged systematically using a single parameter.

We generated the PPs with the OPIUM package.[38] The PPs were generated with LDA and included scalar relativistic corrections. The optimization method of Rappe *et al.*[42] was used to keep the plane-wave energy cutoff below 300 Ry. The PPs include *s, p,* and *d* channels; the *p*-channel was used as local channel. We build the PPs for the doubly (Ca, Sr and Ba) and triply (Y and La) ionized atoms. We tested the PPs by performing DMC calculations of the first and second ionization potential (FIP and SIP) of Ca, Sr, Ba, Sc, Y and La atoms. For Sc, Y and La, we also calculated the third IP (TIP). To calculate the IPs, an imaginary time step of 0.0025 $Ha^{-1}$ and 8192 walkers were sufficient to converge total energies to within 0.001 eV. Single-particle orbitals were generated within DFT and orthorhombic cells with repeated images separated by more than 1.3 nm to ensure isolation. In the subsequent DMC calculations, fully open boundary conditions were employed. For each ion, the known[43] spin multiplicity was adopted.

### C. Zero-point energy and thermal vibrational effects

A consistent comparison of calculated and experimental structural parameters requires inclusion of zero-point energy (ZPE) and thermal vibrational contributions. Experimental parameters are normally determined at ambient conditions, and room temperature thermal expansion could be significant. We studied thermal vibrational effects within the quasi-harmonic approximation. The thermal contributions to the structural parameters are evaluated by fitting the Helmholtz free energy as function of volume to an equation of state. Harmonic phonons were calculated using density functional theory and the linear response method as implemented in the Vienna Ab-initio Simulation Package (VASP).[44–46] For these calculations, we employed projector augmented wave (PAW)[47,48] ionic potentials; PAW potentials with 6, 10 and 11 valence electrons were used for O and atoms of group IIA and IIIB, respectively. The calculations were performed within PW91[49,50] and with a wavefunction energy cutoff of 700 eV. ZPE and thermal properties were evaluated with Phonopy.[51] Results are shown in **Table 1**. Thermal vibrational contributions for the lattice constants, bulk modulus and bulk modulus's pressure derivative were evaluated at 300 K. For the cohesive energies, only ZPE contribution were considered because the experimental data are already extrapolated at 0 K. The ZPE and temperature vibrational contributions included in **Table 1** were added to the calculated cohesive energies and structural parameters to obtain the values reported in Table 3. We use DFT to calculate these contributions instead of QMC because methods for calculating phonons are not yet established for QMC.

**Table 1.** Zero-point energy (ZPE) and thermal vibrational contributions (at 300 K) for the cohesive energy (CE), lattice constant ($a$), bulk modulus (B) and bulk modulus's pressure derivative (B′) of CaO, SrO, BaO, $Sc_2O_3$, $Y_2O_3$ and $La_2O_3$. These contributions were added to the calculated cohesive energies and structural parameters to obtain the values reported in Table 3.

| Oxides | CE (eV) | $a$ (Å) | B (GPa) | B′ |
|---|---|---|---|---|
| CaO | -0.10 | 0.024 | -2.10 | -0.19 |
| SrO | -0.08 | 0.021 | 0.12 | -0.25 |
| BaO | -0.06 | 0.015 | 2.97 | -0.30 |
| $Sc_2O_3$ | -0.32 | 0.038 | -2.15 | -0.13 |
| $Y_2O_3$ | -0.28 | 0.037 | -2.20 | -0.09 |
| $La_2O_3$ | -0.22 | 0.012 | -0.50 | -0.14 |

## III. RESULTS AND DISCUSSION

### A. Pseudopotential testing: Ionization potentials from DMC

To test our PPs, we first studied the sensitivity of the IPs to the choice of nodes (LDA or Heyd-Scuseria-Ernzerhof (HSE06)[52] hybrid functional) and the use of the locality approximation or the T-moves approach in DMC. The IPs evaluated within the locality approximation and the T-moves approaches are similar. The larger difference between the two approaches was found for the FIP of the Lanthanum and Barium atoms, where the FIP evaluated with the locality approximation are 0.05(1) eV closer to experimental values than with the T-moves. The IPs evaluated with single particle orbitals generated with LDA and HSE06 are similar. The only exception is the Lanthanum atom, where the FIP and SIP evaluated with LDA orbitals are, respectively, 0.17(1) and 0.14(1) eV closer to the experimental values than the results with HSE06 orbitals. The DMC total energy of La$^{+1}$ is 0.129(4) eV lower when evaluated with HSE06 instead of LDA orbitals. For all other systems, the DMC total energies evaluated with the two sets of single particle orbitals are similar within an error of 0.04 eV.

**Table 2.** First and second Ionization Potential (IP) of Ca, Sr, Ba, Sc, Y and La calculated with DMC. Results are also included for the third IP of Sc, Y and La. Experimental values[53] are included for comparison.

| Atom | First IP | | Second IP | | Third IP | |
|---|---|---|---|---|---|---|
| | DMC | Exp. | DMC | Exp. | DMC | Exp. |
| Ca | 5.930(5) | 6.113 | 11.858(5) | 11.872 | | |
| Sr | 5.610(7) | 5.695 | 11.022(5) | 11.030 | | |
| Ba | 5.114(4) | 5.212 | 10.001(3) | 10.004 | | |
| Sc | 6.499(6) | 6.561 | 12.585(5) | 12.800 | 24.624(4) | 24.757 |
| Y | 6.056(6) | 6.217 | 12.235(6) | 12.224 | 20.500(4) | 20.524 |
| La | 5.449(5) | 5.577 | 11.086(4) | 11.185 | 19.156(3) | 19.177 |

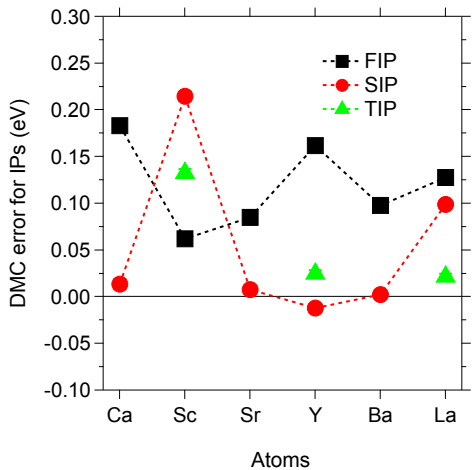

**Figure 1**. Deviation between DMC and experiments for the first (FIP) and second (SIP) Ionization Potentials of Ca, Sc, Sr, Y, Ba, and La. Results are shown also for the third IP (TIP) of Sc, Y, and La. The statistical uncertainty in DMC is smaller than the symbol size.

The DMC calculated IPs are included in **Table 2** along with the corresponding experimental[43] data. The DMC results in **Table 2** were evaluated with the T-moves approach and single particle orbitals generated with LDA. The deviations of the DMC IPs from the experimental values are summarized in Figure 1. The mean absolute error (MAE: $\frac{1}{N}\sum_i^N[|IP_i^{cal} - IP_i^{exp}|]$) across all atoms (Ca, Sr, Ba, Sc, Y and La) is below 0.13 eV for the first (FIP), second (SIP) and third (TIP) ionization potentials: 0.120(2), 0.058(2) and 0.060(2) eV, respectively. The corresponding mean error (ME: $\frac{1}{N}\sum_i^N[IP_i^{cal} - IP_i^{exp}]$) is equal to the MAE for the FIP and TIP, indicating a systematic underestimation of the FIP and TIP. There are not systematic errors for the SIP, the ME is only 0.054(2) eV. Only for the Sc atom, the error in the SIP and TIP slightly deviates from the already small MAE (Figure 1).

Our DMC results for the IPs of Ca, Sr, Ba and La agree well with available all-electron relativistic couple-cluster (RCC) results. For instance, the FIP of Sr,[54] Ba[55] and La[56] calculated with RCC is 5.698, 5.327 and 5.582 eV, respectively. For Ca[57] and Ba,[55] the SIP have been reported at 11.872 and 10.028 eV, respectively. Overall, the RCC IPs are closer to experimental values than our DMC, but the differences between the two methods are of the order of 0.1 eV. One exception is Ba, where the FIP evaluated with RCC deviates from the experimental reference by 0.115 eV while the error in our DMC is -0.098(4) eV. It is also interesting to compare our DMC results with previous DMC calculations. The only relevant calculations we found are the multi-determinant nonrelativistic all-electron DMC calculations[58] for the FIP of Ca and Sc, i.e., 6.04(2) eV and 6.40(2) eV, respectively. Even though a direct comparison of our results with these calculations is not fully justified due to differences in methodologies, the results are in reasonable agreement.

### B. Preliminaries for oxides and expected level of accuracy

As pointed out previously, DMC requires various approximations to be practical:[1] *i)* PP approximation, *ii)* FN approximation, *iii)* short time-approximation and *iv)* supercell approximation. The errors coming from approximations *i* and *ii* are classified as uncontrolled because they cannot be systematically reduced,[19] while errors from approximations *iii* and *iv*, on the other hand, are classified as controlled as they can be systematically reduced. Here we estimate these errors in our DMC calculations of oxides of group IIA and IIIB.

*Uncontrolled errors:* Decoupling and individually analyzing these errors requires reference DMC calculations, e.g., all-electron and beyond FN-DMC calculations, that are impractical for the present systems. Recent studies,[18,59] however, suggest that the primary uncontrolled error is introduced by the PP approximations. Our PPs yield an overall error of 0.13 eV for the IPs of atomic Ca, Sr, Ba, Sc, Y and La. Similarly, the overall error in the cohesive energy of the corresponding oxides is 0.18 eV (see Sec. III.B). It is possible to reduce these errors by employing smaller core PPs, but such PPs are impractical for condensed matter calculations. The errors may also be reduced further by including higher angular momentum channel in the PPs as done in Ref. 60 Nevertheless, we have not tested PPs with higher angular momentum channel (e.g., *f*-channels) because the overall agreement of the calculated properties with experimental values is already excellent. Errors from the FN approximation (FN errors) can be explored employing single-particle orbitals generated with different DFT approximations. As shown previously, the DMC total energies evaluated with LDA and HSE single particle orbitals are similar within an error of 0.04 eV for most atoms of group IIA and IIIB. In our DMC calculations of ZnO,[12] we found that neither the total energy of Zn and O atoms nor the energy of bulk ZnO is significantly affected when using single-particle orbitals from LDA and HSE. On the other hand, our preliminary studies of Co and Fe atoms and bulk CoO and FeO show that total energies of both atom and bulk systems are sensitive to the single-particle orbitals. These results suggest that the FN error in the energy of an atom is an indicator of the degree of this error in solids. Therefore, because we do not find the atoms to have significant FN errors, we do not expect significant FN errors in oxides of group IIA and IIIB due to our choice of LDA orbitals.

*Controlled errors:* Errors from approximation *iv* (time-step errors) can be controlled by simply reducing the DMC time-step. In the present calculations, we employ a DMC time-step of 0.0025 Ha$^{-1}$, which is expected[3] to yield time-step errors within the statistical uncertainty of the DMC results, of the order of 0.05 eV. Errors from approximation *v* (FS-errors) can be classified as one- and two-body FS-errors.[27,61–63] One-body FS-errors were corrected with twist averaged boundary conditions on a 2×2×2 supercell grid. The oxides of group IIA and IIIB that we considered in the present work are insulating materials,[64,65] and 8 twists are expected to be enough to account for one-body FS effects[26] in the employed simulation cells. Calculations within LDA showed that total energies obtained with our simulation cells and a 2×2×2 k-point grid are converged to 1 meV/formula unit (f.u.) with respect to a 6×6×6 k-point grid. Two-body FS-errors are normally corrected by performing calculations with increasing supercell models and extrapolating to infinite volume. This approach increases the computational cost by at least a factor of 3. Alternatively, there are various corrections techniques[27,61–63] that can be employed to remove these FS-errors and avoid the additional cost of extrapolating. We employed the KZK method.[27] In our previous DMC calculations of the EOS ZnO, we compared the KZK and extrapolation techniques. Our results show that the KZK method removes the bulk of the two-body FS-error but a residual FS-error of the order of 0.1 eV/f.u. remains present for supercell models of 32 atoms. We performed similar test calculations for SrO and BaO, and estimated a residual FS-error 0.07 eV for the supercell model of 54 atoms.

## C. Structural properties of binary oxides of groups IIA and IIIB

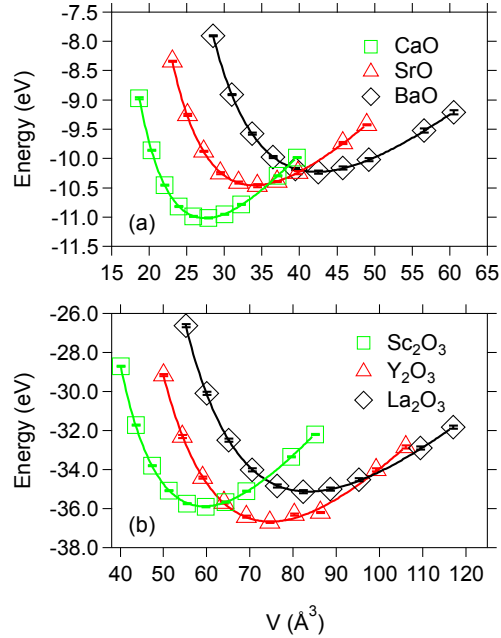

**Figure 2.** DMC energy versus volume for binary oxides of (a) group IIA (CaO, SrO and BaO) and (b) group IIIB ($Sc_2O_3$, $Y_2O_3$ and $La_2O_3$) together with fitted equations of state (Murnaghan). Energies are per formula unit and relative to the Ca, Sr, Ba, Sc, Y, La and O atoms. The statistical uncertainty in DMC is smaller than the symbol size.

Curves of DMC energy versus volume are shown in Figure 2 for binary oxides of groups IIA (CaO, SrO and BaO) and IIIB ($Sc_2O_3$, $Y_2O_3$ and $La_2O_3$). The data was fitted to the Murnaghan equation of state to determine the structural parameters (cohesive energy, lattice constants, bulk modulus and bulk modulus's pressure derivative). In addition to the DMC calculations, we also evaluated the structural parameters of the binary oxides with LDA, PW91 and HSE06[52] hybrid functionals calculations. The LDA and PW91 calculations were performed with Quantum ESPRESSO,[29] employing our PPs and the supercell models described above. The HSE06 calculations were performed with VASP,[44–46] PAW[47,48] potentials and an energy cutoff of 550 eV. We note that LDA and PW91 have been extensively applied to study the structural properties of binary oxides of group IIA and IIIB (see Refs. 66 and 67 and references in there). Our results are in general agreement with these previous calculations. We do not include any detailed comparisons with previous LDA and PW91 calculations for the sake of clarity and brevity. The derived parameters from our DMC, LDA, PW91 and HSE06 calculations are included in **Table 3**. Results in **Table 3** include ZPE and thermal vibrational effects as evaluated within PW91 (see Sec. II.C). To quantify the accuracy of our DMC, LDA, PW91 and HSE06 calculations, we compare the derived parameters with experimental measurements. The errors of the derived parameters are shown graphically in Figure 3. Statistics of the errors are included in Table 4.

**Table 3.** Cohesive energy and structural parameters (lattice constants, bulk modulus and bulk modulus's pressure derivative) of CaO, SrO, BaO, $Sc_2O_3$, $Y_2O_3$, and $La_2O_3$ evaluated with DMC, LDA and PW91.[a] Available experimental results are included for comparison. The statistical uncertainty in DMC is provided in parentheses.

| Oxides | DMC | LDA | PW91 | HSE | Expt. |
|---|---|---|---|---|---|
| Cohesive energy (eV) | | | | | |
| CaO | 10.92(1) | 12.57 | 10.68 | 10.34 | 10.95 |
| SrO | 10.38(2) | 11.82 | 9.96 | 9.66 | 10.37 |
| BaO | 10.16(2) | 11.56 | 9.73 | 9.35 | 10.09 |
| $Sc_2O_3$ | 35.55(3) | 40.37 | 34.06 | 33.55 | 35.15 |
| $Y_2O_3$ | 36.37(4) | 40.66 | 34.77 | 34.05 | 36.05 |
| $La_2O_3$ | 34.89(8) | 39.10 | 32.91 | 33.39 | 35.14 |
| Lattice constant $a$ (Å) | | | | | |
| CaO | 4.811(1) | 4.734 | 4.836 | 4.840 | 4.811[32] |
| SrO | 5.160(1) | 5.087 | 5.194 | 5.193 | 5.16[33] |
| BaO | 5.546(1) | 5.478 | 5.589 | 5.596 | 5.54[34] |
| $Sc_2O_3$ | 9.865(1) | 9.749 | 9.910 | 9.900 | 9.85[35] |
| $Y_2O_3$ | 10.648(7) | 10.498 | 10.659 | 10.680 | 10.60[35] |
| $La_2O_3$ | 3.966(1) | 3.923 | 3.984 | 3.933 | 3.934[36] |
| Bulk modulus B (GPa) | | | | | |
| CaO | 110.62(7) | 119.54 | 101.66 | 106.63 | 111(1)[68] |
| SrO | 92.29(8) | 98.36 | 83.56 | 88.00 | 91,[33] 90.6[69] |
| BaO | 76.41(9) | 82.36 | 70.34 | 73.79 | 74.06[70] |
| $Sc_2O_3$ | 183.65(9) | 185.75 | 165.54 | 177.03 | 154(5)[71] |
| $Y_2O_3$ | 152.30(4) | 152.15 | 136.38 | 143.59 | 176[72] |
| $La_2O_3$ | 133.34(7) | 138.22 | 123.19 | 133.07 | 113.0[72] |
| Bulk modulus's pressure derivative B′ | | | | | |
| CaO | 4.06(1) | 4.23 | 3.94 | 3.90 | 4.2(2)[68] |
| SrO | 4.06(1) | 4.25 | 3.97 | 3.97 | 4.3[33] |
| BaO | 4.07(1) | 4.21 | 3.99 | 3.98 | 5.67[70] |
| $Sc_2O_3$ | 3.75(1) | 3.97 | 3.78 | 3.74 | 7[71] |
| $Y_2O_3$ | 3.41(9) | 4.06 | 3.88 | 3.87 | 5.1[72] |
| $La_2O_3$ | 3.87(1) | 4.03 | 4.03 | 3.92 | 6.0[72] |

[a] The calculated cohesive energies were corrected for zero point energy (ZPE), and the structural parameters were corrected for thermal vibrational contributions at 300 K (**Table 1**). The experimental[73,74] cohesive energies were estimated combining the measured (extrapolated at 0 K) dissociation energy of $O_2$, the cohesive energies of the metal crystals and the enthalpies of formation of the binary oxides.

**Table 4.** Comparison of calculated and measured cohesive energy, lattice constant, bulk module and its pressure derivative for binary oxides of group IIA and IIIB. The error statistics are the mean error (ME: $\frac{1}{6}\sum_{i=1}^{6}[x_i^{cal} - x_i^{exp}]$), the mean absolute error (MAE: $\frac{1}{6}\sum_{i=1}^{6}[|x_i^{cal} - x_i^{exp}|]$), the mean relative error (MRE: $\frac{1}{6}\sum_{i=1}^{6}[\frac{x_i^{cal}-x_i^{exp}}{x_i^{exp}}\times 100]$), and the mean absolute relative error (MARE: $\frac{1}{6}\sum_{i=1}^{6}[|\frac{x_i^{cal}-x_i^{exp}}{x_i^{exp}}|\times 100]$). The statistical uncertainty in DMC is provided in parenthesis.

|  | DMC | LDA | PW91 | HSE |
| --- | --- | --- | --- | --- |
| **Cohesive energy** | | | | |
| ME (eV) | 0.09(2) | 3.06 | -0.94 | -1.23 |
| MAE (eV) | 0.18(2) | 3.06 | 0.94 | 1.23 |
| MRE (%) | 0.30(7) | 13.72 | -3.82 | -5.81 |
| MARE (%) | 0.64(7) | 13.72 | 3.82 | 5.81 |
| **Lattice constant *a*** | | | | |
| ME (Å) | 0.017(1) | -0.07 | 0.05 | 0.04 |
| MAE (Å) | 0.017(1) | 0.07 | 0.05 | 0.04 |
| MRE (%) | 0.26(1) | -1.07 | 0.75 | 0.59 |
| MARE (%) | 0.26(1) | 1.07 | 0.75 | 0.59 |
| **Bulk modulus B (GPa)** | | | | |
| ME (GPa) | 5.43(7) | 9.55 | -6.40 | 0.51 |
| MAE (GPa) | 13.45(7) | 17.50 | 13.64 | 13.86 |
| MRE (%) | 5.35(5) | 9.40 | -4.60 | 1.12 |
| MARE (%) | 9.95(5) | 13.91 | 10.10 | 9.79 |
| **Bulk modulus's pressure derivative B′** | | | | |
| ME | -1.51(2) | -1.07 | -1.45 | -1.48 |
| MAE | 1.51(2) | 1.21 | 1.45 | 1.48 |
| MRE (%) | -25.4(3) | -20.46 | -24.37 | -25.00 |
| MARE (%) | 25.4(3) | 20.71 | 24.37 | 25.00 |

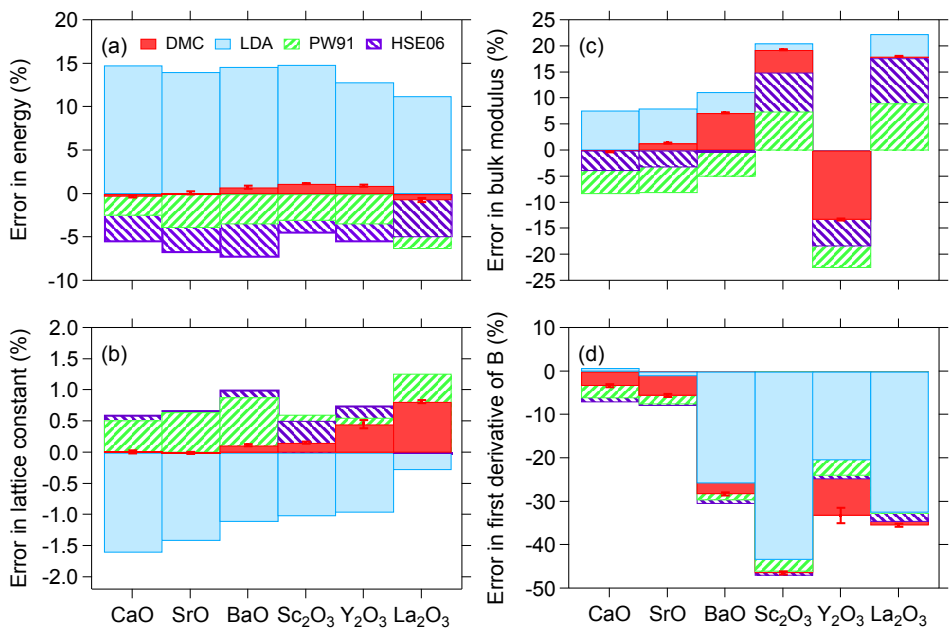

**Figure 3.** Comparison (relative error) of calculated and measured (a) cohesive energy, (b) lattice constant, (c) bulk modulus and (d) its pressure derivative for binary oxides of group IIA and IIIB. Standard deviation error bars show statistical uncertainties in the DMC data.

Overall, DMC outperforms PW91, HSE06 and LDA for the cohesive energy of the binary oxides; the MAE are 0.18(2), 0.94, 1.23 and 3.06 eV for DMC, PW91, HSE06 and LDA, respectively (Table 4). The corresponding ME are 0.09(2), -0.94, -1.23 and 3.06 eV. As expected,[75] the cohesive energy of the oxides are systematically underestimated and overestimated in PW91 and LDA, respectively (Figure 3). The HSE06 cohesive energies are similar to those obtained with PW91. This similarity has been previously reported for the formation energy of MgO, CaO and various oxides of Titanium.[76] In the case of DMC, the cohesive energy is generally overestimated. The exceptions are CaO and $La_2O_3$, where DMC underestimates the energies by 0.03(1) and 0.25(8) eV, respectively (**Table 3**). DMC yields the cohesive energy of oxides of group IIA in remarkable agreement with experimental values; the deviations are below 0.07(2) eV. For oxides of group IIIB, the DMC errors are larger, reaching a value of 0.40(3) eV in $Sc_2O_3$. The differing accuracy for the two types of oxides is expected. Oxides of group IIIB are naturally more complex because they involve bonding with *d*-orbitals. Moreover, accurate cohesive energies require correct description of the solid and isolated atomic system. In our DMC calculations, the singlet state of atoms of group IIA are likely better described than the doublet state of atoms of group IIIB, for which a single Slater determinant times a Jastrow factor may not be sufficient. These uncertainties, which ultimately are the result of deviations of the trial wave function from the exact ground state, come from the use of nonlocal pseudopotentials and the fixed-node approximation.

In the case of the lattice constant, DMC also perform better than PW91, LDA and HSE06. The mean absolute relative errors (MARE) for the lattice constant are 0.26(1)%, 0.75%, 1.07% and 0.59% for DMC, PW91, LDA and HSE06, respectively. Similar to the cohesive energy, the DMC lattice constants of the oxides of group IIA are in excellent agreement with measurements (Figure 3). The agreement worsens for oxides of group IIB. In the case of $La_2O_3$, the DMC error for the lattice constant is much larger than the rest at 0.81(9)%. The stable phase of $La_2O_3$ has a hexagonal structure, and we evaluated the EOS approximately by fixing the *c/a* ratio to the experimental[36] value. Evaluating the EOS under such approximation can yield the lattice constant 0.4% larger than with full cell relaxation (as tested within PW91). The large error for DMC in the lattice constant of $La_2O_3$ can partially be the result of our approximation to evaluate the EOS. This could also explain that the PW91 and LDA errors for the lattice constant of $La_2O_3$ are outside the corresponding deviation found for $Sc_2O_3$ and $Y_2O_3$ (Figure 3).

For the bulk modulus and its pressure derivative, the overall agreement with experimental values is relatively poor for the all methods. The MARE values are above 9% and 20% for the bulk modulus and its pressure derivative, respectively (Table 4). Inspection of Figure 3 shows, however, that the high MARE values for the bulk modulus are clearly caused by the oxides of group IIIB. MARE values for the bulk modulus evaluated including only oxides of group IIA are 3.00(5)% 7.20% and 9.00% for DMC, PW91 and LDA, respectively. DMC performs reasonably well for the bulk modulus of CaO, SrO and BaO. Although all these methods have systematic errors, the consistently large deviations from experimental values suggest that both the predictions and experimental data need to be carefully re-examined. A similar conclusion can be reached for the bulk modulus's pressure derivative of BaO and the oxides of group IIIB (Figure 3).

For these oxides, the MARE values for the bulk modulus's pressure derivative are over 20% in all calculations. However, in the case of CaO and SrO, all methods reproduce the experimental bulk modulus's pressure derivative within 8%.

Our DMC results are consistent with the benchmark calculations presented in Ref. 3 for the volumes and bulk moduli of many different type of solids. For instance, the MARE value for the DMC equilibrium volumes of the oxides that we considered is below 1.0%, in proximity to the value of 2.20(7)% reported in Ref. 3. For the for bulk moduli, our overall MARE value for DMC, 9.95(5)%, is larger than the value of 5.03(48)% reported in Ref. 3. However, due to the limited number of experimental measurements for the bulk moduli, it would be helpful for additional experiments to be conducted. Our results corroborate the findings of Ref. 3 showing that DMC is highly accurate in describing the structural properties of a broad range of solids, including oxides. The results supports that structural properties of solid are rather insensitive to the nodal surface,[3] but they also highlight the need for further benchmarking, understanding, and control of PPs related errors for heavy elements.

Our calculations for oxides of groups IIA and IIIB are one of the first efforts to systematically assess how accurate standard DMC methods can predict the cohesive energies of solids. We have employed relatively small supercell models in our calculations and approximate finite-size error corrections. Without any doubt, greater accuracies are possible with smaller core pseudopotentials, better trial wavefunctions, and a more careful treatment of finite-size error as already pointed out in Ref. 3. Nevertheless, our results show that highly accurate cohesive energies can be obtained for simple oxides already with standard DMC methods.

## IV. SUMMARY

In summary, we have used standard DMC methods and the local and semi-local DFT approximations to calculate the cohesive energy and the structural parameters of the binary oxides CaO, SrO, BaO, $Sc_2O_3$, $Y_2O_3$ and $La_2O_3$. DMC outperforms local and semi-local DFT approximations in describing the cohesive energies and structural parameters of these binary oxides. DMC yields cohesive energies with a mean absolute error of 0.18(2) eV, while in local, semi-local and hybrid DFT approximations it is 3.06, 0.94 and 1.23 eV, respectively. For lattice constants, the mean absolute errors in DMC, local, semi-local and hybrid DFT approximations, are 0.017(1), 0.07, 0.05 and 0.04 Å, respectively. The results clearly show that DMC is an attractive alternative to DFT. The calculations presented in this paper, not only are a test of the accuracy of DMC method, but also a validation of the accuracy of the pseudopotentials used for a key set of group IIA and IIIB elements. In combination with transition metal pseudopotentials tested in Ref. 77 these new pseudopotentials allow the calculation of a large number of highly correlated oxides of broad interest.

ACKNOWLEDGMENT

The work was supported by the Materials Sciences & Engineering Division of the Office of Basic Energy Sciences, U.S. Department of Energy. Paul R. C. Kent was supported by the Scientific User Facilities Division, Office of Basic Energy Sciences, U.S. Department of Energy. Computational resources were provided by the Oak Ridge Leadership Computing Facility at the Oak Ridge National Laboratory, supported by the Office of Science of the U.S. Department of Energy under Contract No. DE-AC05-00OR22725.